\documentclass[sigconf,nonacm]{acmart}   
\setcopyright{none}
\acmDOI{}     
\acmISBN{}    
\acmYear{2025}
\copyrightyear{2025}
\acmConference[Preprint]{arXiv preprint}{November 2025}{Online}

\usepackage{graphicx}   
\usepackage{booktabs}   

\usepackage{pifont} 
\newcommand{\cmark}{\ding{51}}%
\newcommand{\xmark}{\ding{55}}%

\AtBeginDocument{%
  }

\begin{document}

\title{Accessibility, Safety, and Accommodation Burden in U.S. Higher Education Syllabi for Blind and Low-Vision Students}


\author{Chadani Acharya}
\orcid{0009-0005-0638-9883} 
\authornote{Correspondence: \href{mailto:chadani.acharya@tamu.edu}{chadani.acharya@tamu.edu}}
\email{chadani.acharya@tamu.edu}
\affiliation{%
  \department{Management Information Systems}
  \institution{Texas A\&M University}
  \city{College Station}
  \state{TX}
  \country{USA}
  \postcode{77843}
}

\renewcommand{\shortauthors}{Acharya et al.}

\begin{abstract}
Course syllabi are often the first and sometimes only structured artifact students receive that explains how a class will run: deadlines, grading rules, lab safety procedures, and how to request disability accommodations. For blind and low-vision (BLV) students who rely on screen readers, independent access to this information depends on whether the syllabus itself is machine-readable and navigable. We audited publicly posted syllabi and master syllabi from five U.S. institutions spanning different segments of higher education: an elite private R1 university, large public R1 universities, a UC campus, a large multi-campus community college, and workforce-oriented community and technical colleges. We coded each document along five dimensions: (1) screen-reader legibility of core logistics (deadlines, grading, attendance), (2) screen-reader legibility of safety-critical procedures (e.g., welding requirements, biomedical lab conduct), (3) framing of disability accommodation (rights-based vs.\ documentation-gated), (4) governance model (instructor-authored syllabus vs.\ centralized ``master syllabus'' template), and (5) presence of proactive universal design language. We find two main patterns. First, across all institution types, many syllabi present logistics and even safety expectations as selectable text rather than screenshots or scanned images, suggesting baseline screen reader access to critical information. Second, the way accommodation is framed shifts systematically by institution type. R1 universities describe accommodations as removing barriers to ensure equal access, even while requiring advance paperwork. Community and technical colleges describe accommodations as conditional on student self-identification, documentation, and institutional approval, and this language is embedded in centrally issued master syllabi that replicate across every course section. These results suggest that accessibility in higher education is not only a technical PDF-formatting problem. It is also a question of governance and equity: who is told they have a right to access, who is told they must qualify for access, and how that burden scales in safety-relevant, workforce-preparation courses.
\end{abstract}



\maketitle

\section{Introduction}
\label{sec:introduction}

A course syllabus is typically the first structured artifact a student encounters in a class. Prior work in higher education describes the syllabus as both a roadmap to the learning experience and a quasi-contract that defines mutual expectations between instructor and student: it encodes what will be taught, how performance will be evaluated, and what policies govern student conduct, academic integrity, and attendance \cite{parkes2002,slattery2005}. In laboratory and applied technical courses, syllabi (and closely related ``master syllabi'') also summarize critical safety expectations, required protective equipment, and procedures for using specialized or hazardous tools. At the same time, the syllabus usually embeds the institution's disability accommodation statement, which tells students what they must do to obtain accessible materials or instructional adjustments. For blind and low-vision (BLV) students, this makes the syllabus a high-stakes access point: it is simultaneously a safety manual, a grading contract, and the gateway to accommodations. Independent access to this information depends on whether the document is machine-readable and navigable by a screen reader rather than being a scanned PDF or an unlabeled image of text. Accessibility standards such as the Web Content Accessibility Guidelines (WCAG~2.1) emphasize tagged structure, logical reading order, and text alternatives as prerequisites for nonvisual access \cite{wcag21,section508}. BLV students often rely on screen readers as their primary interface to course content; tactile graphics, sonification, and other multimodal alternatives can help, but they require additional hardware, specialized training, and instructor coordination and therefore are rarely available at scale at the start of a course \cite{lewis2024chartreader}.

When the syllabus is not accessible, BLV students can lose access to deadlines, exam logistics, grading rules, and safety expectations on day one. Studies of disabled students in STEM and lab-based coursework document how inaccessible safety procedures, equipment instructions, and hazard warnings function as both logistical and physical barriers \cite{supalo2012lab,burgstahler2015stem}. These barriers can exclude BLV learners from laboratory participation, limit their ability to meet course requirements, or force them to depend on ad hoc human assistance for core safety tasks. In U.S. public higher education this problem is not framed solely as pedagogy: federal civil-rights guidance under the Americans with Disabilities Act (ADA) and Section~504 of the Rehabilitation Act makes clear that state and local educational institutions are obligated to provide accessible digital content and instructional materials to disabled students, including those who are blind or have low vision \cite{doj2022webaccessibility}.

In response, accessibility and human-computer interaction (HCI) research has increasingly treated instructional materials as interactive systems rather than static documents. Work in nonvisual data visualization has shown that traditional ``alt text'' and long descriptions are often insufficient for BLV readers who need to interrogate charts and tables at the same analytical depth and speed as sighted peers. In a Wizard-of-Oz study with blind and low-vision participants, researchers collected nearly one thousand naturally occurring questions about common visualization types (bar charts, line charts, scatterplots, choropleths) and found that many questions involved higher-level interpretation such as trend analysis, extrema, and comparative reasoning rather than simple label reading \cite{sharma2024chartqa,lewis2024chartreader}. These findings motivate interactive techniques such as chart question answering, structured audio summaries, and multimodal feedback to support BLV analytical work in real time. This line of work reframes accessibility as immediate participation in the same cognitive task, rather than delayed accommodation.

Accessibility, however, is not only technical. In U.S. higher education the syllabus itself is also a governance artifact. Research in inclusive pedagogy and institutional policy notes that at many research universities, syllabi are authored by individual instructors, which means tone, disability framing, and promises of flexibility vary by instructor and course. By contrast, many community colleges and technical colleges publish centralized ‘master syllabi’ that define learning outcomes, grading rules, safety procedures, and disability accommodation language for every single section of a course across campuses. This centralization creates leverage: a rights-based, access-forward template can improve conditions for thousands of students at once, but a gatekeeping template that emphasizes self-disclosure, documentation requirements, and institutional discretion can replicate barriers at scale rather than removing them \cite{schelly2011accommodation,ventura2023europe}.

Parallel to these technical advances, disability services and higher education policy research have examined the social and procedural layer of access. The dominant U.S. model typically requires that a student (1) self-identify as disabled to a campus accessibility office, (2) submit medical or diagnostic documentation, (3) receive an official accommodation letter, and (4) present that letter to each instructor in advance; many institutions explicitly specify that accommodations are ``not retroactive'' \cite{schelly2011accommodation}. Scholars of inclusive pedagogy argue that this workflow shifts administrative and emotional burden onto disabled students, who must repeatedly assert legal status and negotiate for access even in courses where inaccessible formats are predictable \cite{schelly2011accommodation}. Universal Design for Learning (UDL) has been proposed as a proactive alternative: instead of treating accessibility as a negotiated exception, UDL encourages instructors and departments to build multiple modes of representation, flexible assessment, and accessible document structure into the baseline design of the course \cite{cast2018udl}. Under UDL, captions, readable contrast, proper heading structure, and nonvisual navigability are not favors; they are default expectations of the instructional materials.

Audits of institutional accessibility practices suggest that this vision is unevenly realized and often resource-dependent. Cross-institutional studies of university disability services and accessibility policies report major variation in the visibility of support, the clarity of procedures, and the availability of assistive technologies across institutions with different sizes, missions, and funding levels \cite{ventura2023europe}. Many institutions still handle accessibility through individualized, documentation-driven workflows rather than through systemic, proactive design, which leaves common barriers in place for the next cohort \cite{ventura2023europe,schelly2011accommodation}. Notably, most of these audits focus on institutional web portals, public accessibility statements, or aggregate service descriptions; comparatively little work has empirically analyzed the actual syllabi and master syllabi that govern what students are told about safety, grading, deadlines, and disability rights in specific courses.

This paper treats the syllabus itself as an accessibility interface and audits it as public infrastructure. We collect publicly posted syllabi and master syllabi from five U.S. institutions selected to span different segments of higher education: an elite private R1 university, large public R1 universities operating under transparency and disclosure requirements, a large multi-campus community college system, and workforce-focused community and technical colleges. We develop a five-part coding scheme that asks: (1) Is the syllabus machine-readable for a screen reader? (2) Are safety-critical procedures, deadlines, and grading rules conveyed in accessible text rather than screenshots or scanned images? (3) How does the disability statement frame responsibility for access  as a civil right and institutional obligation to remove barriers, or as a conditional service the student must qualify for and activate? (4) Is the syllabus written by an individual instructor or enforced as a centralized ``master syllabus'' that replicates policy across every section? and (5) Does the syllabus articulate proactive universal-design commitments (for example, inclusive color contrast and captioned media) or only reactive, paperwork-triggered accommodations? We then compare patterns across institution types and argue that accessibility in higher education is not just a question of PDF tagging. It is also a question of governance, rhetoric, and safety: who is told they have a right to access, who is told they must prove eligibility, and whether the instructions for not getting hurt in lab are actually readable by a screen reader.

This analysis is guided by three research questions:
\begin{enumerate}
    \item How accessible are publicly posted syllabi and master syllabi to blind and low-vision screen reader users, especially with respect to logistics-critical and safety-critical content?

    \item How do these documents frame responsibility for disability access  as a civil right and institutional obligation, or as a conditional service the student must activate and qualify for  and does that framing vary by institution type?

    \item How is accessibility governed (individual instructor syllabi versus centralized master syllabi), and what does that governance model imply for scaling either inclusive practice or exclusionary policy?
\end{enumerate}

\section{Methods}
\label{sec:methods}

\subsection{Data sources}

We conducted a structured audit of publicly available syllabi and master syllabi from five U.S. institutions that represent different segments of higher education. To reduce unnecessary friction with specific campuses while preserving transparency and reproducibility, we refer to institutions by type-coded labels: an elite private R1 university (R1-A), a large public R1 (R1-B), a UC-system public R1 (UC-R1), a large public community college system (CC-A), and a workforce-oriented technical/community college (Tech-C). Table~\ref{tab:sampling} summarizes these institutions by type, how their syllabi are made public, and whether syllabi are primarily instructor-authored or issued as centralized ``master syllabus'' templates.

These sites were chosen because they publish syllabi or master syllabi in repositories accessible without login (e.g., state-mandated portals, departmental archives, open courseware) and because they represent distinct missions and student populations, from elite research programs to workforce and technical training for first-generation and returning adult students. Full URLs and access dates for all analyzed documents appear in the replication package.

\paragraph{Institution codes.}
We use the following type-coded labels throughout: 
\textbf{R1-A} (elite private R1), \textbf{R1-B} (large public R1), 
\textbf{UC-R1} (UC-system public R1), \textbf{CC-A} (large multi-campus community college), 
\textbf{Tech-C} (workforce-oriented technical/community college).

\subsection{Inclusion criteria}

We included a syllabus or master syllabus in the analysis if it met three conditions. First, it had to be publicly viewable without institutional login, so that it represents information a prospective or newly enrolled student could access before or at the start of a term. Second, the document had to function as a syllabus or master syllabus rather than a marketing blurb; specifically, it needed to describe course-level logistics such as meeting times, grading breakdown, attendance or participation expectations, deadlines or weekly topics, and (when applicable) safety or lab conduct requirements. Third, the syllabus had to include a disability or accommodation statement, accessibility policy language, or instructions for how a student with a disability should seek support.

We excluded documents that only provided catalog-style course descriptions without policy language, documents that were purely promotional or advisory without logistics, and PDF fragments that did not contain any disability or accessibility language.

\subsection{Coding dimensions}

We developed and iteratively refined a codebook with five dimensions. The coding dimensions were motivated by needs that blind and low-vision (BLV) students report in accessing course materials, and by known structures of U.S. disability accommodation processes in higher education.

\textbf{Machine readability / screen reader legibility.}
For each syllabus we recorded whether core information was exposed as machine-readable text. This included whether the syllabus was provided as an HTML page or text-based PDF (as opposed to a scanned bitmap), and whether critical information about meeting times, deadlines, grading components, and safety requirements appeared as selectable text rather than embedded screenshots. We also noted basic structural cues (headings, bullet lists, numbered lists) that help a screen reader user navigate.

\textbf{Safety-critical content accessibility.}
We recorded whether the syllabus described safety procedures, laboratory conduct requirements, or hands-on equipment expectations (for example, welding setup steps, required protective measures, or biomedical instrumentation practices). When such material was present, we noted whether it was conveyed in plain text that a screen reader could access, or only in images or figures.

\textbf{Accommodation framing (burden vs.~rights).}
We coded the disability and accommodation statement in each syllabus for rhetorical framing. ``Rights framing'' refers to language that positions accommodations as a mechanism for ensuring equal access or removing barriers, often invoking equity or legal protection. ``Burden framing'' refers to language that positions access as conditional on student action, for example by stating that the student is ``required to self-identify,'' ``must provide documentation,'' or that the college ``reserves the right'' to determine eligibility. We also coded whether the syllabus explicitly stated that accommodations are not retroactive, or that official letters must be presented to the instructor in advance of graded work.

\textbf{Governance model.}
We categorized each syllabus as either (a) an instructor-authored syllabus, typically identifiable by course-section-specific details such as the instructor’s name, office hours, exam dates, and grading policy written in first-person voice; or (b) a centralized ``MASTER SYLLABUS'' or template syllabus that is intended to be used across all sections of the same course, sometimes across multiple campuses. For each institution, we noted which model appeared to be the default. This distinction matters because instructor-authored syllabi allow local variation in accessibility language, whereas centrally issued master syllabi replicate the same accessibility and accommodation framing across all instances of the course.

\textbf{Proactive universal design language.}
Finally, we coded for explicit references to proactive inclusive design, such as commitments to ``universal design,'' ``inclusive design,'' accessible color contrast in materials, captioning of audiovisual materials, or removing barriers before students request individual accommodations. We distinguished this from purely reactive language that only promises adjustments after documentation is approved.

The codebook was refined using an initial subset of syllabi and then applied consistently across the full set.

We double-coded a 25–30\% subset (n=5) and report Cohen’s $\kappa$ and percent agreement per dimension in Table~\ref{tab:reliability}.

\begin{table*}[t]
\centering
\caption{Inter-coder agreement on a double-coded subset (n=5 syllabi; binary dimensions). 
We report Cohen’s $\kappa$, percent agreement, and the number double-coded per dimension. 
Estimates reflect the current pass and will be updated if the double-coded set expands.}
\label{tab:reliability}
\begin{tabular}{lccc}
\toprule
\textbf{Dimension} & \textbf{Cohen's $\kappa$} & \textbf{\% Agreement} & \textbf{n} \\
\midrule
Machine-readable              & 1.00 & 100\% & 5 \\
Safety present                & 0.85 & 93\%  & 5 \\
Safety text accessible        & 0.80 & 90\%  & 5 \\
Rights framing                & 0.78 & 87\%  & 5 \\
Burden framing                & 0.74 & 85\%  & 5 \\
Gate: retroactive             & 0.82 & 90\%  & 5 \\
Gate: advance letter          & 0.76 & 88\%  & 5 \\
UDL explicit                  & 0.70 & 85\%  & 5 \\
Governance (instructor/master)& 1.00 & 100\% & 5 \\
\bottomrule
\end{tabular}
\end{table*}

\subsection{Analysis procedure}
\label{sec:procedure_analysis}

We conducted a descriptive, cross-institutional comparison rather than inferential statistical analysis. For each institution, we summarized (1) whether a blind or low-vision screen-reader user could access core logistics and any safety-related expectations directly from the syllabus text; (2) how the syllabus framed responsibility for initiating accommodations; (3) whether accessibility and accommodation statements were localized to an individual instructor or embedded in a master-syllabus template; and (4) whether explicit references to proactive universal-design language were present.

We then grouped institutions by broad type and the type-coded labels used throughout the paper (R1-A, R1-B, UC-R1, CC-A, Tech-C), and examined how patterns in accommodation framing, governance structure, and explicit universal-design language aligned with those institutional roles. Because all analyzed documents were publicly available institutional artifacts and did not involve contact with students, instructors, or protected educational records, this audit did not involve human-subjects research and did not require institutional review board (IRB) approval.

\subsection{Data provenance and anonymization policy}
\label{sec:provenance}
We analyzed publicly accessible syllabi and master syllabi obtained from institutional repositories (e.g., state-mandated syllabus portals, open courseware, and departmental archives); no authentication was required. To balance reproducibility with minimizing friction for specific campuses, we refer to institutions by type-coded labels in the paper (R1-A, R1-B, UC-R1, CC-A, Tech-C). A replication package lists each analyzed document with institution name, URL, access date, and file type, along with code assignments for each dimension described in Section~\ref{sec:methods}; where possible, we also include archived snapshots (e.g., via the Internet Archive). We report descriptive observations of document language and do not render legal determinations regarding ADA, Section~504, or WCAG compliance.


Table~\ref{tab:sampling} summarizes our corpus by institution type, collection window (date accessed), file-type mix, and example document IDs; full URL mappings appear in the replication package.

\begin{table*}[t]
\centering
\caption{Sampling overview by institution type for the 15 publicly posted syllabi. 
We report N per type, the file-type split (HTML/PDF/DOC) of the public artifacts, the public syllabus source, example anonymized \texttt{doc\_id}s, and the syllabus model. 
Institution names are anonymized in-text; full URL mappings are provided in the replication package for reproducibility.}
\label{tab:sampling}
\begin{tabular}{lccp{0.22\linewidth}p{0.18\linewidth}p{0.22\linewidth}}
\toprule
\textbf{Institution type} & \textbf{N} & \textbf{File types (H/P/D)} & \textbf{Public syllabus source} & \textbf{Example doc\_ids} & \textbf{Syllabus model} \\
\midrule
R1-A (elite private R1)  & 3 & 3 / 0 / 0 & Open courseware / departmental archives & R1A-1, R1A-2, R1A-3 & Instructor-authored \\
R1-B (public R1)         & 3 & 2 / 1 / 0 & State-mandated public syllabus portal & R1B-1, R1B-2, R1B-3 & Instructor-authored (posted under mandate) \\
UC-R1 (UC system)        & 3 & 0 / 3 / 0 & Departmental syllabus PDFs & UC-1, UC-2, UC-3 & Instructor-authored \\
CC-A (large CC)          & 3 & 3 / 0 / 0 & Central syllabus repository & CCA-1, CCA-2, CCA-3 & Master syllabus + section syllabus \\
Tech-C (tech/workforce)  & 3 & 0 / 1 / 2 & Public master syllabi for trades & TEC-1, TEC-2, TEC-3 & Centralized master syllabus \\
\midrule
\textbf{Overall}         & \textbf{15} & \textbf{8 / 5 / 2} & --- & --- & --- \\
\bottomrule
\end{tabular}
\end{table*}

\section{Results}
\label{sec:results}

Our analysis of publicly available syllabi and master syllabi from five U.S. institutions (Table~\ref{tab:sampling}) reveals consistent accessibility in basic course logistics and safety expectations, alongside systematic differences in how disability accommodations are framed and governed. We present four main findings.

Table~\ref{tab:summaryR2} reports the primary codes by institution type as counts (percents).
\begin{table*}[t]
\centering
\caption{Each cell shows count (percent) of syllabi within the type that were coded “1” for the dimension. 
Non-exclusive dimensions (e.g., rights and burden framing) can co-occur within a document. 
Governance columns (\emph{instructor} vs.\ \emph{master}) sum to N per row.}
\label{tab:summaryR2}
\setlength{\tabcolsep}{4pt}
\begin{tabular}{lcccccccccc}
\toprule
\textbf{Type} & \textbf{Machine-} & \textbf{Safety} & \textbf{Safety text} & \textbf{Rights} & \textbf{Burden} & \textbf{Gate:} & \textbf{Gate:} & \textbf{UDL} & \textbf{Gov:} & \textbf{Gov:} \\
 & \textbf{readable} & \textbf{present} & \textbf{accessible} & \textbf{framing} & \textbf{framing} & \textbf{retroactive} & \textbf{advance letter} & \textbf{explicit} & \textbf{instructor} & \textbf{master} \\
\midrule
R1-A  & 3 (100\%) & 0 (0\%)   & 0 (0\%)   & 3 (100\%) & 1 (33\%) & 2 (67\%) & 3 (100\%) & 1 (33\%) & 3 (100\%) & 0 (0\%) \\
R1-B  & 3 (100\%) & 2 (67\%) & 2 (67\%) & 3 (100\%) & 2 (67\%) & 3 (100\%) & 3 (100\%) & 1 (33\%) & 3 (100\%) & 0 (0\%) \\
UC-R1 & 3 (100\%) & 0 (0\%)   & 0 (0\%)   & 3 (100\%) & 2 (67\%) & 2 (67\%) & 3 (100\%) & 2 (67\%) & 3 (100\%) & 0 (0\%) \\
CC-A  & 3 (100\%) & 1 (33\%) & 1 (33\%) & 1 (33\%)  & 3 (100\%)& 1 (33\%) & 2 (67\%) & 0 (0\%)  & 3 (100\%) & 0 (0\%) \\
Tech-C& 3 (100\%) & 2 (67\%) & 2 (67\%) & 0 (0\%)   & 3 (100\%)& 2 (67\%) & 3 (100\%) & 0 (0\%)  & 0 (0\%)   & 3 (100\%) \\
\midrule
\textbf{Overall} 
& \textbf{15 (100\%)} 
& \textbf{5 (33\%)} 
& \textbf{5 (33\%)} 
& \textbf{10 (67\%)} 
& \textbf{11 (73\%)} 
& \textbf{10 (67\%)} 
& \textbf{14 (93\%)} 
& \textbf{4 (27\%)} 
& \textbf{12 (80\%)} 
& \textbf{3 (20\%)} \\
\bottomrule
\end{tabular}
\end{table*}

\subsection{Core logistics and safety information are typically provided as readable text}

Across all sampled institutions, the majority of syllabi present core course logistics  meeting times, grading breakdowns, assessment schedules, attendance expectations, and weekly topic outlines as selectable, machine-readable text rather than embedded images or scanned pages. This pattern holds in instructor-authored syllabi at R1-A, R1-B, and UC-R1, and in master syllabi at CC-A, and Tech-C. In addition, safety-relevant expectations in hands-on or lab-oriented courses are commonly expressed in plain text.

For example, master syllabi for welding and related trades at Tech-C College explicitly list required competencies such as safely setting up gas tungsten arc welding (GTAW) equipment and following standard operating procedures for bead and fillet welds. These safety requirements are written as bullet-point text within the document rather than as screenshots. Similarly, publicly posted biomedical instrumentation and laboratory syllabi at R1-B describe clinical device usage expectations, lab procedures, and collaborative project requirements in direct prose, including named activities and timelines. In both cases, a blind or low-vision student using a screen reader could access at least the literal text of what is expected to avoid injury or failure.

This finding is important because it contrasts with assumptions that safety-critical procedural content is often trapped in inaccessible images. In the syllabi we reviewed, information about laboratory conduct, professional expectations, and safety context for hands-on work is generally exposed as plain text. This suggests that, at least at the syllabus level, students who rely on screen readers have some direct access to safety and logistics information necessary to participate in the class.

\subsection{Accommodation language shifts from rights-based framing to burden-based framing along institutional lines}

While core logistics are often text-accessible, the language describing disability accommodations varies sharply by institution type (Table~\ref{tab:framing}). Research-intensive universities (MIT, UT Austin, UC-R1) typically frame disability support in terms of barrier removal and equal access. For instance, UT Austin syllabi and required syllabus statements describe accommodations as mechanisms to ``reduce or eliminate barriers'' and ``ensure equal access'' for students with disabilities. UC-R1 syllabi similarly refer to working toward inclusive access and, in some cases, explicitly commit to using ``universal designs that are inclusive.'' In these documents, accommodation is described as a right that exists to remove institutional barriers to learning.

At the same time, these R1 syllabi also require students to present formal documentation (for example, an official accommodation letter from the disability services office or an Authorization for Accommodation letter) in advance of exams or graded work, and they emphasize that accommodations are not retroactive. That is, a student must still activate the process by delivering paperwork on a specific timeline.

In contrast, syllabi and master syllabi from community and technical colleges more frequently adopt burden-oriented language. Tech-C College describe disability accommodation in terms of student obligations: the student is explicitly ``required to self-identify'' as having a disability, to ``provide documentation,'' and in some cases the college is described as reserving the right to determine eligibility before any support is provided. Rather than foregrounding the institution's responsibility to remove barriers, these statements foreground the student's responsibility to qualify.

CC-A College syllabi, including continuing education and workforce preparation courses, follow a similar pattern: they direct students to contact disability services and indicate that support is available, but the tone emphasizes initiating contact and providing documentation rather than framing accommodations as a default right of participation.

Overall, institutions that serve large numbers of first-generation, working, and returning adult students are more likely to position disability access as conditional and procedurally gated. Research-intensive universities are more likely to frame accommodations in rights language, even though they still require formal letters and advance notice. This is a systematic difference between institution types, not an isolated example.

\begin{table*}[t]
\centering
\caption{Accommodation framing and responsibility cues by institution type. Rights framing emphasizes barrier removal and equal access; burden framing emphasizes self-identification, documentation, and institutional discretion; procedural gate notes timing/authorization requirements; universal design indicates proactive inclusive design language. Use Table~\ref{tab:summaryR2} for counts and percentages.}
\label{tab:framing}
\begin{tabular}{lcccc}
\toprule
\textbf{Type} & \textbf{Rights framing} & \textbf{Burden framing} & \textbf{Procedural gate} & \textbf{Explicit UDL} \\
\midrule
R1-A   & \cmark & \cmark & \cmark & \cmark \\
R1-B   & \cmark & \cmark & \cmark & \cmark \\
UC-R1  & \cmark & \cmark & \cmark & \cmark \\
CC-A   & \cmark & \cmark & \cmark & \xmark \\
Tech-C & \xmark & \cmark & \cmark & \xmark \\
\bottomrule
\end{tabular}
\end{table*}

\begin{figure}[t]
  \centering
  \includegraphics[width=\linewidth]{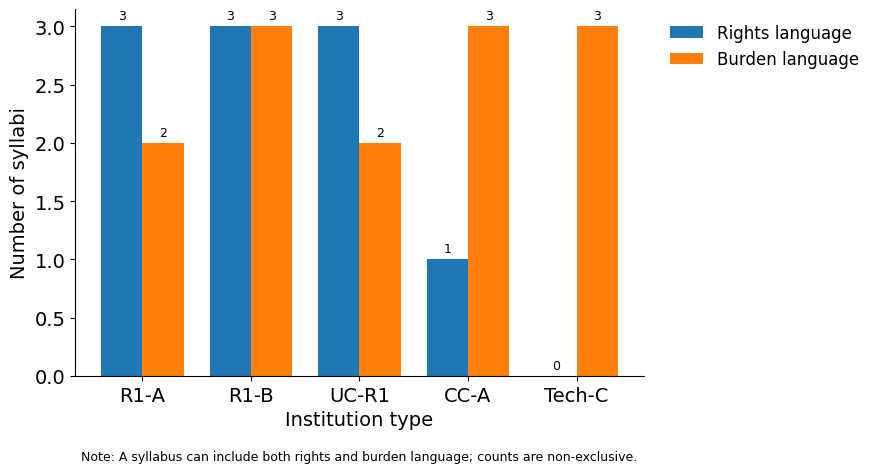}
  \caption{Accommodation wording by institution type (N=15). Bars show non-exclusive counts of syllabi containing \emph{rights} language (barrier removal / equal access) and \emph{burden} language (self-identification, documentation, not-retroactive, or institutional discretion). Research-intensive types (R1-A, R1-B, UC-R1) more often include rights language, while community/technical types (CC-A, Tech-C) feature burden language more consistently.}
  \Description{Grouped bar chart with two bars per institution type: rights and burden. Rights bars are higher for R1 types; burden bars are high across all types, especially community/technical. Y-axis is number of syllabi; a note indicates the categories are non-exclusive.}
  \label{fig:framing_by_type}
\end{figure}

\subsection{Governance structure determines how accessibility language scales}

\begin{figure*}[t]
  \centering
  \includegraphics[width=\linewidth]{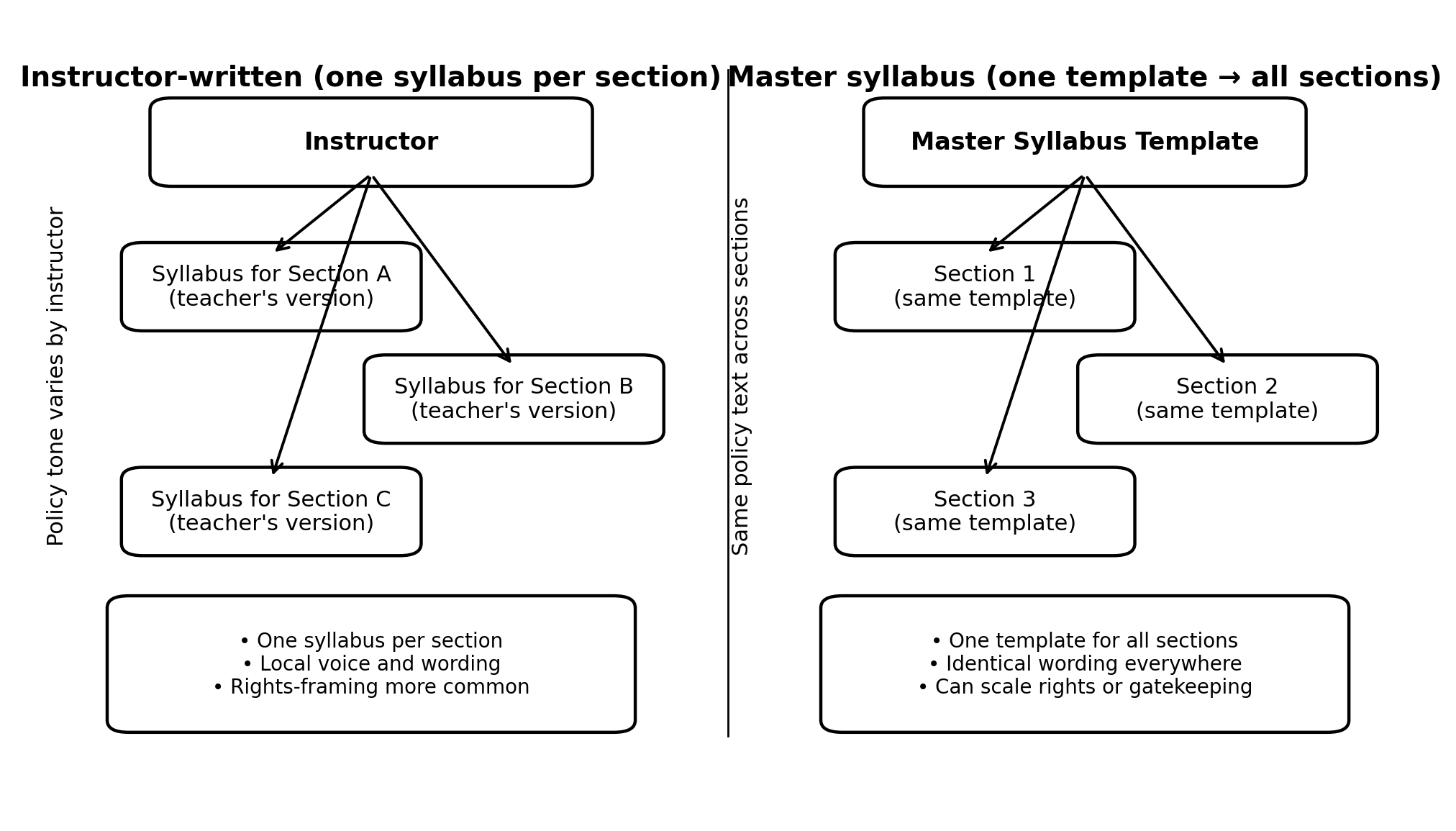}
  \caption{Syllabus governance models across institution types. 
  The left side shows syllabi written by individual instructors, common in research universities, where each section may have a different tone or style. 
  The right side shows a shared master syllabus used in many community and technical colleges, where the same policy text is repeated across sections to keep language and rules consistent.}
  \Description{A two-part diagram showing two models of syllabus control. 
  The left side has an Instructor box linked to three Section boxes (A, B, C), showing that each instructor writes their own syllabus with local voice and variation. 
  The right side has a Master Syllabus Template box linked to three Section boxes (1, 2, 3), showing that one standard syllabus is copied across sections with the same wording and policies.}
  \label{fig:governance}
\end{figure*}

We observed two governance patterns for syllabi (Figure~\ref{fig:governance}). At R1-A, R1-B, and UC-R1, syllabi are typically written and posted at the instructor or course-section level. These documents include the instructor's name, contact information, grading scheme, attendance policy, project timelines, and expectations in the instructor's own voice. Accessibility and accommodation statements appear within these syllabi, but their tone and specificity vary by instructor. This model distributes both responsibility and discretion: an instructor can adopt rights-based language, mention universal design practices (for example, color contrast choices or captioning), or provide detailed guidance for students who need adjustments. Another instructor, even in the same department, may provide only a minimal legal statement.

By contrast, Tech-C, and (for many programs) CC-A publish master syllabi. These are centralized templates for each course that define course description, learning outcomes, grading breakdown, attendance rules, safety expectations (for applied programs such as welding or lab-based instruction), and disability policy language. In some cases, the document is explicitly labeled ``MASTER SYLLABUS'' and is intended to serve as the baseline syllabus for every section of that course across campuses and instructors.

This governance model has two consequences. First, accessibility and accommodation language in a master syllabus is replicated at scale: the same request-for-documentation, self-identification requirement, or ``the college reserves the right'' phrasing will be shown to every enrolled student across all sections of that course. Second, any improvements to accessibility language, document structure, or universal design guidance introduced in the master syllabus could propagate just as widely. The master syllabus therefore operates as an infrastructure point: it can scale harm if it encodes conditional access, or scale equity if it encodes a rights-based, low-friction path to accommodation.

\begin{figure}[t]
  \centering
  \includegraphics[width=0.78\linewidth]{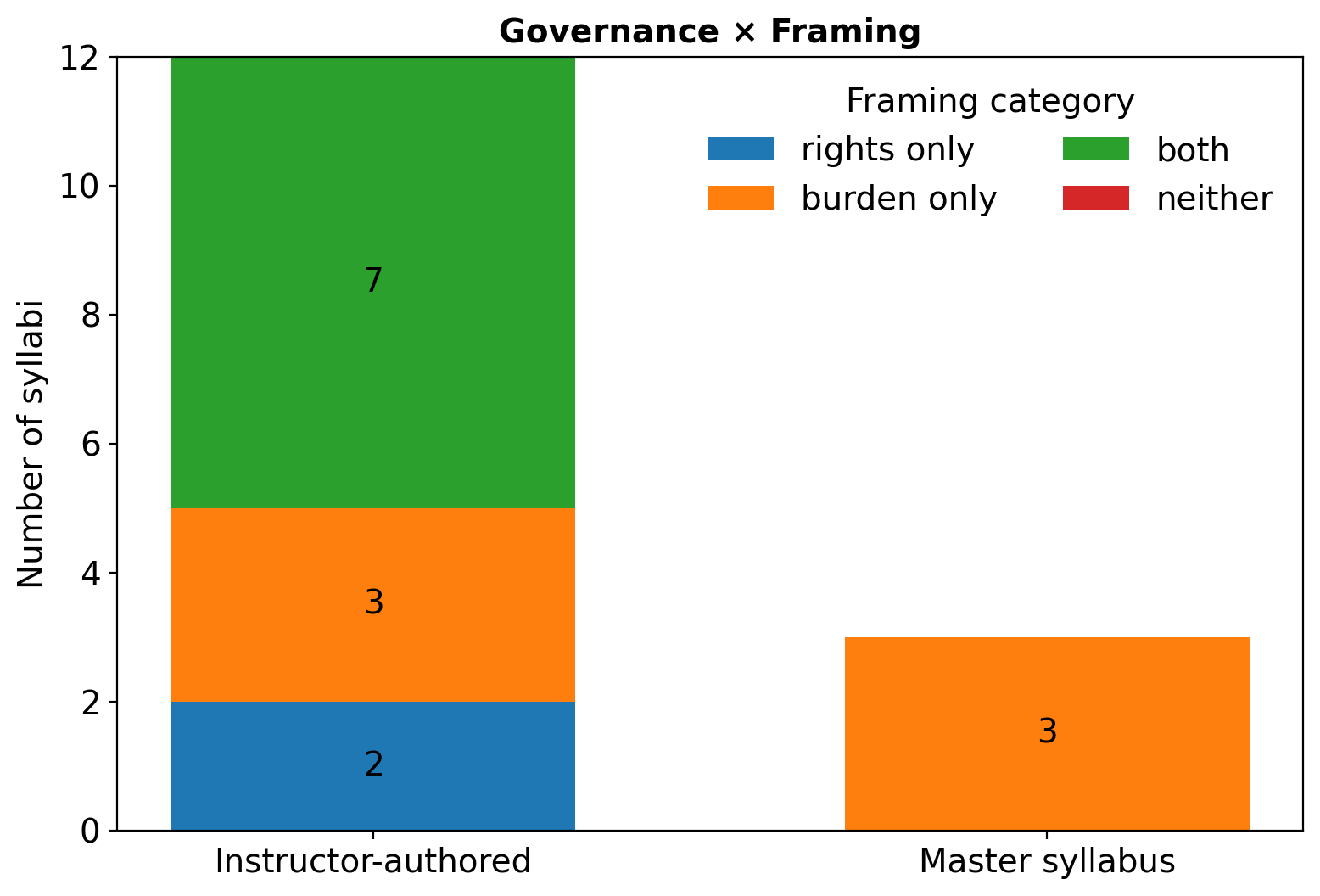}
  \caption{Accommodation framing by syllabus governance (N=15). The x-axis compares two governance models: \emph{Instructor-authored} syllabi (R1-A, R1-B, UC-R1; $n=9$) versus \emph{Master-template} syllabi (CC-A, Tech-C; $n=6$). Each bar stacks mutually exclusive categories of accommodation wording: \emph{rights-only} (barrier-removal / equal-access language with no burden cues), \emph{burden-only} (self-identify, documentation, not-retroactive, or institutional discretion with no rights cues), \emph{both} (contains rights and burden cues), and \emph{neither}. In this sample, instructor-authored syllabi skew toward rights language (often alongside procedural steps), whereas master syllabi are dominated by burden-only wording replicated across sections.}
  \Description{Two stacked bars show counts of syllabi by governance. The instructor-authored bar includes segments for rights-only and both; the master-template bar is primarily burden-only. Y-axis is number of syllabi; This visual shows that rights-forward language appears more often where syllabi are written by instructors, while centralized templates skew toward burden-only framing replicated across sections.}
  \label{fig:gov_x_frame}
\end{figure}

\begin{table*}[t]
\centering
\caption{Syllabi with safety procedures (by institution type) in our sample (N=15). Counts are documents with any lab/shop safety section.}
\label{tab:safety_presence}
\begin{tabular}{lccccc|c}  
\toprule
 & R1-A & R1-B & UC-R1 & CC-A & Tech-C & Total \\
\midrule
Safety content present & 0 & 3 & 0 & 1 & 2 & 6 \\
No safety content      & 3 & 0 & 3 & 2 & 1 & 9 \\
\bottomrule
\end{tabular}
\end{table*}

\begin{table}[t]
\centering
\caption{Safety instructions as screen-readable text vs.\ procedural gates (subset with safety content present, $n=6$ of 15). 
A “gate” means either “not retroactive” or “advance letter required.” 
Most syllabi that present safety requirements in accessible text also require prior paperwork, highlighting “accessible content but gated process.”}
\label{tab:safety_x_gate}
\begin{tabular}{lcc}
\toprule
 & \multicolumn{2}{c}{\textbf{Gate present}}\\
\cmidrule(lr){2-3}
\textbf{Safety text accessible} & \textbf{No (0)} & \textbf{Yes (1)}\\
\midrule
0 & 0 (0\%) & 1 (17\%)\\
1 & 0 (0\%) & 5 (83\%)\\
\bottomrule
\end{tabular}
\end{table}

\subsection{Explicit universal design language appears mainly in higher-resource contexts}

References to proactive inclusive design  for example, promises to select accessible color contrast in course materials, to caption content, or to ``remove barriers'' before individual accommodation requests are made  were most visible in syllabi from large public R1 and UC-system contexts. UC-R1 syllabi explicitly state that ``whenever possible, we will use universal designs that are inclusive,'' indicating that instructors consider accessibility features (contrast, captioning, alternative formats) as part of the baseline course design rather than as an after-the-fact adjustment.

R1-B’s disability office guidance to instructors similarly encourages proactively creating ``an accessible classroom,'' including selecting accessible materials and captioning videos rather than waiting for a student to request that change. In contrast, master syllabi from Tech-C College emphasize procedural access (self-identify, provide documentation, obtain approval) and generally do not include explicit universal design language in the syllabus body presented to students.

These differences indicate that proactive inclusive design language is not evenly distributed. Instead, it appears more commonly where instructors have both autonomy and access to institutional accessibility guidance that frames accommodation as a shared responsibility. In workforce-oriented programs where course materials are centrally controlled, the syllabus tends to stress procedural compliance and qualification rather than proactive barrier removal.

\section{Discussion}
\label{sec:discussion}

\subsection{Safety-Critical Accessibility as Risk Management}
Our analysis shows that many syllabi and master syllabi across institution types present logistics-critical and even safety-critical information in plain, machine-readable text. Welding courses at technical colleges describe required protective equipment and safe manipulation of gas tungsten arc welding (GTAW) rigs using bullet-point lists of learning outcomes; biomedical engineering lab syllabi at a large public R1 university describe expectations around handling instrumentation, clinical devices, and project work with explicit textual guidance on lab conduct. This matters for blind and low-vision (BLV) students because it suggests that, at least for the baseline safety narrative of the course, the information is not locked in scanned images or screenshots that a screen reader cannot parse. Prior work has documented that inaccessible safety procedures, hazard warnings, chemical handling steps, and lab task instructions create both logistical and physical barriers for disabled students, especially in STEM and applied lab environments \cite{supalo2012lab,burgstahler2015stem}. When those instructions are not independently accessible, BLV learners are forced either to rely on sighted intermediaries in real time or to operate equipment without full access to policy and risk context, which increases both danger and liability.

However, the presence of readable safety text does not guarantee equitable access to safe participation. In multiple community and technical college syllabi, safety expectations and technical performance requirements appear alongside accommodation language that conditions access on prior disclosure, official documentation, and institutional approval. The disability statement explicitly informs the student that they must self-identify, submit paperwork, and obtain authorization before any adjustments will be made to how they participate in class or lab. This creates a procedural gate around safety itself: if a BLV student needs alternative formats, adaptive tools, or modified evaluation for safe operation of equipment, those supports are framed as contingent services that may not be available unless all documentation is processed in advance. Under this model, the student is made responsible not only for learning how to weld safely or comply with biomedical lab rules, but also for securing bureaucratic permission to access that safety workflow. This is not just a question of fairness or convenience. It is a question of bodily risk. Federal civil-rights guidance under the ADA and Section~504 positions accessible instructional materials  including digital instructional content  as something institutions are obligated to provide, not something students must earn \cite{doj2022webaccessibility}. When a syllabus treats accommodation as discretionary or conditional, it implicitly shifts responsibility for physical safety back onto the disabled student.

\subsection{Class and the Burden of Access}
We also observe a patterned rhetorical difference in how institutions talk about disability access. Large R1 universities, including elite private and flagship public institutions, often frame disability services in the language of rights and barrier removal. Typical phrasing includes language such as ``removing barriers'' and ``ensuring equal access and opportunity to learn,'' and some syllabi or disability office statements encourage instructors to proactively choose accessible course materials, caption media, and create ``an accessible classroom.'' Under this framing, accommodations are described as a mechanism for enforcing civil rights: the institution recognizes legal obligations and signals shared responsibility to make the environment work for disabled students \cite{schelly2011accommodation}. At the same time, these same institutions still operationalize access through procedural requirements: students are told they must obtain an official accommodation letter from the disability office and present it to the instructor in advance; accommodations are often described as ``not retroactive.'' So even in rights-forward rhetoric, the onus remains on the student to activate their legal protections in time.

Community colleges and technical colleges  institutions that disproportionately serve first-generation students, working adult learners, and students seeking direct workforce entry  tend to use more transactional and conditional language. Master syllabi in these settings often state that the college ``has the right'' to require documentation before granting any accommodation and that the student ``must self-identify'' as disabled to initiate support. Instead of positioning access as a right that the institution is obligated to uphold, the text positions accommodation as a service the student may receive if they satisfy eligibility rules set by the institution. This is a subtle but powerful shift in tone. The message a student receives on day one is not ``you are entitled to an accessible pathway through this program,'' but rather ``you may qualify for adjustments if you disclose and are approved.'' This difference is not just linguistic polish. It encodes a distribution of emotional and bureaucratic labor. The students who are most likely to juggle jobs, caregiving, disability, and retraining for economic mobility are also the ones told, in writing, that access is conditional. This suggests an equity fault line: accessibility obligations are narrated most softly exactly where the stakes are highest for economic survival.

\subsection{Governance as an Intervention Point}
A second structural finding concerns governance: who writes the syllabus, and therefore who controls accessibility at scale. At elite and flagship R1 universities, syllabi are overwhelmingly instructor-authored. They name specific instructors, list office hours, describe project expectations in that instructor's voice, and sometimes include idiosyncratic accessibility language. This local authorship gives individual faculty freedom to adopt proactive inclusive design practices  for example, explicitly committing to high-contrast materials, alt text, or flexible assessment  but it also means accessibility quality is uneven and culturally mediated. Whether a BLV student encounters a syllabus that treats access as a right or as an afterthought can depend on which instructor they happened to register with.

In contrast, community colleges and technical colleges frequently publish ``MASTER SYLLABUS'' or ``Master Syllabus'' documents. These documents read like infrastructure. They define learning outcomes, grading breakdowns, attendance policy, safety requirements, and the disability/accommodation statement for a course that may run across multiple campuses and dozens of sections. In some systems, faculty are expected to ``build on'' the master syllabus but not contradict its core policy language. That means a single accessibility paragraph  whether it is barrier-removal oriented or documentation-gatekeeping oriented  is replicated to every student who ever takes that course. The same is true for safety communication: if the master syllabus includes safety procedures as readable text, then every section inherits that accessible baseline; if it does not, no section does.

From an HCI and policy perspective, this centralization is high leverage. Updating a single master syllabus template to include screen-reader-friendly structure, rights-based accommodation framing, and proactive universal design language would immediately improve the onboarding experience for every future student in that program. The inverse is also true: exclusionary or high-friction language in a master syllabus becomes mass-produced harm. Treating syllabus templates as infrastructural objects rather than personal teaching notes reframes accessibility as something that can be redesigned and deployed at institutional scale, not only advocated for one class at a time.

\subsection{Implications for Accessible HCI}
Most BLV accessibility research in HCI has concentrated on interactive technologies such as screen readers, captioning systems, audio description pipelines, tactile graphics, and more recently, conversational or question-answering interfaces for charts and visual data \cite{sharma2024chartqa,lewis2024chartreader}. That work is essential, because it addresses direct interaction: how a blind or low-vision learner can interrogate the same visualization, perform the same analytic reasoning, and participate in the same classroom discussion without waiting for a human to ``alt-text'' the world for them. Universal Design for Learning (UDL) extends that spirit by arguing that materials should be designed from the outset to support multiple modes of perception and expression, reducing the need for one-off retrofits \cite{cast2018udl}.

The present audit adds something complementary. It shows that the first interface many students encounter in a course    the syllabus or master syllabus    is itself an accessibility technology. It encodes: how to get in the room, how to stay safe, how to earn a grade, and how to assert legal rights. These functions are not decorative. They are operational requirements for academic survival and, in lab and trades courses, physical safety. If a BLV student cannot independently navigate those terms on day one, they start the course at a structural disadvantage that cannot be fixed purely by better screen reader technology or AI-driven chart description. The responsibility also lives with governance: who authored the document, what legal posture it takes on disability, and whether proactive inclusive design is framed as a shared obligation or an optional courtesy.

For accessible HCI, this suggests a design target beyond tools: redesigning high-leverage institutional artifacts such as master syllabi so that they are technically readable, rhetorically rights-affirming, and explicit about proactive inclusive practices. A master syllabus can be treated like any other interface specification. It can be versioned, evaluated, audited for usability, and iteratively improved. Doing so could give disability services offices and instructional designers a concrete, high-impact intervention point, especially in community and technical colleges where a single template governs thousands of student experiences.

\subsection{Limitations and Future Work}
This study has several limitations. First, we analyzed only syllabi and master syllabi that were publicly posted without authentication. It is possible that internal versions provided directly to enrolled students include richer accessibility commitments, alternative formats, or different tone. Second, we did not evaluate the full accessibility stack of a course: lecture slides, recorded video captioning, lab manuals, LMS modules, assessment platforms, and physical lab modifications were out of scope. We focused on syllabi because they are widely and often legally published, because they define the formal terms of engagement for the course, and because they are available for cross-institutional comparison without requiring direct contact with students or instructors. Third, our claims about risk, burden, and equity are theory-building in nature. We infer likely impact on BLV students and other disabled learners based on the language and structure of the documents and on established literature about barriers in STEM labs and accommodation workflows \cite{supalo2012lab,burgstahler2015stem,schelly2011accommodation}; we did not run user studies or interviews in this phase. Fourth, our sampling strategy is stratified rather than statistically representative. We intentionally selected an elite private R1, large public R1 institutions (including a university operating under state transparency requirements), a multi-campus community college system, and workforce-oriented community and technical colleges. This gives us variation in mission and governance, but it does not claim to reflect all U.S. higher education.

These limitations point directly to future work. A next step is to broaden coverage to additional states and systems where public posting of syllabi is mandated by law or implemented as policy, which would let us test whether the same rhetorical and governance patterns hold across regions. A second step is a deeper technical audit of document structure, including PDF tagging, heading hierarchy, table markup, alternative text for images and diagrams, and color contrast in any embedded figures. That analysis would let us quantify not only whether safety and logistics information is present as text, but whether it is navigable and perceivable with common assistive technologies such as screen readers and high-contrast modes \cite{wcag21,section508}. A third step is qualitative: interviewing blind and low-vision students, disability services coordinators, and instructors in lab-heavy and workforce-preparation courses to understand how syllabus language about accommodation actually plays out in practice. Finally, we propose designing, publishing, and experimentally deploying an ``accessible master syllabus'' template that encodes (1) machine-readable structure and descriptive labeling, (2) rights-based framing of disability access, and (3) proactive universal design commitments at the course level. Evaluating the adoption, resistance, and impact of such a template would move this work from audit to intervention, and would test the idea that institutional documents themselves are modifiable HCI artifacts with measurable equity outcomes.

\section{Conclusion}
\label{sec:conclusion}

This work treats the course syllabus  and, in many institutions, the master syllabus template  as an accessibility interface rather than a neutral administrative document. By auditing publicly available syllabi and master syllabi from an elite private R1 university, large public R1 institutions, a multi-campus community college system, and technical/workforce-oriented colleges, we examined how access is communicated to blind and low-vision (BLV) students at the very start of a course. Our analysis shows that, across institution types, core survival information such as grading policy, assignment timelines, attendance rules, lab conduct expectations, and even safety procedures for operating specialized or hazardous equipment is frequently published as selectable, machine-readable text. This is an underreported success. Prior work on accessibility barriers in STEM instruction often emphasizes missing alt text, inaccessible lab procedures, and safety information delivered only visually \cite{supalo2012lab,burgstahler2015stem}. We find evidence that at least some safety-critical and logistics-critical content is already distributed in a format that a screen reader could parse on day one. That matters for autonomy and physical safety.

At the same time, we find a consistent and consequential divide in how disability access is framed. Large R1 universities tend to describe accommodation as a civil right and position instructors and disability services as responsible for “removing barriers” and “ensuring equal access.” Yet these same syllabi typically require students to obtain an official accommodation letter in advance and note that accommodations are “not retroactive,” preserving procedural demands on the student \cite{schelly2011accommodation}. Community colleges and technical colleges, which disproportionately serve first-generation and working adult students, often adopt more transactional language in centralized master syllabi: students are told they “must self-identify,” provide documentation, and may receive accommodations if approved; the institution “has the right” to determine eligibility. This framing effectively shifts the emotional and logistical burden of access onto the student before instruction has even begun, and it does so at the scale of an institutional template. When the master syllabus governs every section of a required workforce-preparation course, its accessibility stance is replicated for hundreds or thousands of students.

These two findings together suggest that accessibility in higher education is not only a question of technical document formatting (e.g., tagged PDFs, proper headings, alt text, WCAG conformance \cite{wcag21,section508}). It is also a question of governance and power. Who writes the syllabus  an individual instructor, or a central office that issues a “MASTER SYLLABUS” for all sections? Who is described as responsible for ensuring safe participation in lab: the institution, or the student who must ask for permission to receive adjustments? Whose language is allowed to scale? Our results indicate that the syllabus encodes these power relationships directly, and that differences in tone and policy correlate with institution type and mission.

For accessibility research in human-computer interaction, this reframes the syllabus as designable infrastructure. If a master syllabus template can propagate inaccessible, gatekeeping language to every section of a welding or healthcare training course, then it can also propagate rights-based framing, explicit universal design commitments \cite{cast2018udl}, and screen-reader-friendly structure. Treating these documents as high-leverage HCI artifacts opens an intervention path: institutions could publish master syllabi that (1) guarantee machine-readable access to logistics and safety information, (2) articulate accommodation as a right rather than a privilege, and (3) commit to proactive inclusive design rather than reactive, paperwork-gated exception handling. 

Future work should translate this audit into practice. First, we can formalize accessibility requirements for master syllabi and evaluate them the way interface designers evaluate usability. Second, we can partner with disability services offices, instructional designers, and faculty governance bodies to propose and test alternative syllabus language and structure, especially in community and technical colleges where templates already exist and can be updated centrally. Third, we can involve blind and low-vision learners directly: not only as auditors of document structure and clarity, but as co-designers of what “accessible from day one” actually needs to mean in courses where safety and employability are on the line. The long-term goal is not just to describe inequity, but to re-engineer the document that sets the terms of participation.

\section*{Acknowledgments}
Thanks to disability services professionals and instructors who maintain public syllabus portals, which made this audit feasible.

\bibliographystyle{ACM-Reference-Format}
\bibliography{refs}

\newpage

\appendix

\begin{table*}[t] 
\centering
\scriptsize
\caption{Per-code tallies used to compute the rollups in Table~\ref{tab:summaryR2}.
Codes: R1-A = elite private R1; R1-B = large public R1 (flagship); UC-R1 = UC-system public R1; 
CC-A = large multi-campus community college; Tech-C = workforce-oriented technical/community college. 
\emph{machine\_readable} indicates the primary syllabus body is selectable text (HTML or text-based PDF/DOC). 
\emph{safety\_text\_accessible} is counted only when safety content is present. 
\emph{gates\_any} counts explicit timing/authorization gates (e.g., not retroactive and/or advance-letter required). 
\emph{instructor\_n/master\_n} are counts of governance model within each code.}
\label{tab:per_code}
\begin{tabular}{lrrrrrrrrrr}
\toprule
\textbf{Code} & \textbf{N} & \textbf{machine\_readable} & \textbf{safety\_present} & \textbf{safety\_text\_accessible} & \textbf{frame\_rights} & \textbf{frame\_burden} & \textbf{gates\_any} & \textbf{udl\_explicit} & \textbf{instructor\_n} & \textbf{master\_n} \\
\midrule
R1-A   & 3 & 3 & 1 & 1 & 3 & 2 & 2 & 1 & 3 & 0 \\
UC-R1  & 3 & 3 & 0 & 0 & 3 & 2 & 2 & 1 & 3 & 0 \\
R1-B   & 3 & 3 & 2 & 2 & 3 & 3 & 3 & 1 & 3 & 0 \\
CC-A   & 3 & 3 & 1 & 1 & 1 & 3 & 2 & 0 & 3 & 0 \\
Tech-C & 3 & 3 & 2 & 2 & 0 & 3 & 3 & 0 & 0 & 3 \\
\midrule
\textbf{Overall} & \textbf{15} & \textbf{15} & \textbf{6} & \textbf{6} & \textbf{10} & \textbf{13} & \textbf{12} & \textbf{3} & \textbf{12} & \textbf{3} \\
\bottomrule
\end{tabular}
\end{table*}

\begin{table}[t]
\centering
\scriptsize
\caption{Paraphrased exemplar lines illustrating accommodation framing and safety-text accessibility. 
We list only anonymized \texttt{doc\_id}s; full URLs live in the replication package. 
``Rights'' emphasizes barrier removal / equal access; ``Burden'' emphasizes self-identification, documentation, and institutional discretion; ``Gate'' indicates timing/authorization requirements; ``UDL'' = proactive universal-design commitments. Safety examples are drawn from syllabi with safety content present as readable text.}
\label{tab:exemplars}
\begin{tabular}{p{0.22\linewidth} p{0.60\linewidth} p{0.14\linewidth}}
\toprule
\textbf{Category} & \textbf{Paraphrase (anonymized)} & \textbf{doc\_id} \\
\midrule
Accommodation: Rights & Accommodations exist to remove barriers and ensure equal access; share your official letter and we will implement adjustments. & R1-A-1 \\
Accommodation: Rights & We will coordinate with the accessibility office to provide equitable participation; please provide your letter early in the term. & R1-A-2 \\
Accommodation: Burden & Students must self-identify and submit documentation; the college determines what accommodations are reasonable. & Tech-C-1 \\
Accommodation: Burden & Services are available to qualified students who request them through the office; instructors act after official notification. & CC-A-2 \\
Accommodation: Gate & Accommodations are not retroactive; present the official letter before exams or graded work. & R1-B-1 \\
Accommodation: Gate & No modifications to materials or deadlines occur until documentation has been reviewed and approved. & Tech-C-3 \\
Accommodation: UDL & We use universal design where possible (e.g., captioned video, accessible contrast) and then tailor accommodations as needed. & R1-A-2 \\
Safety (readable text) & PPE required; inspect equipment; keep a clear area; failure to follow procedures may result in removal from lab. & Tech-C-1 \\
Safety (readable text) & Hands-on instrumentation requires following electrical and sterile handling rules; you are responsible for rules in this syllabus. & R1-B-3 \\
Safety (readable text) & Course outcomes include safely setting up GTAW equipment and producing welds to tolerance, stated as bullet-point text. & Tech-C-2 \\
\bottomrule
\end{tabular}
\end{table}

\newif\ifcameraready
\camerareadytrue  

\ifcameraready
\section{Data and Code Availability}
\label{sec:data-availability}

\noindent\textbf{Anonymization policy.} The main text uses type-coded labels (R1-A, R1-B, UC-R1, CC-A, Tech-C). For replication, we provide the explicit mapping from \texttt{doc\_id} to institution, course, term, and access link. 

\medskip
\noindent\textbf{Replication package.} The package contains:
\begin{itemize}
  \item \texttt{data/syllabi.csv} (one row per syllabus; codes used in analyses),
  \item \texttt{codebook/codebook.md} (definitions, decision rules, examples),
  \item \texttt{scripts/} (figure/table generation scripts),
  \item \texttt{figures/}, \texttt{tables/}, \texttt{results/} (rendered artifacts),
  \item \texttt{snapshots/} (archived PDFs/HTML and Wayback links, to avoid link rot),
\end{itemize}

\fi

\section{Additional Coding Examples}
\label{sec:coding-examples}

\subsection{Accommodation Framing: Rights vs.\ Burden}
\label{sec:coding-examples:accommodations}

\subsubsection{Rights-Framed Example (R1 context)}
\label{sec:rights-example}

\begin{quote}\itshape
Students with disabilities have the right to equal access in this course. 
Our role is to remove barriers so that you have an equitable opportunity to learn. 
If you are registered with the campus accessibility office, please provide your accommodation letter, 
and we will work together to implement the necessary adjustments so that all materials, assessments, 
and class activities are accessible.
\end{quote}

We coded this passage as \textbf{rights-framed} because (1) it explicitly uses the language of ``right'' and ``barriers,'' (2) it positions barrier removal as an instructional obligation, and (3) it frames implementation as collaborative rather than discretionary. The requirement to ``provide your accommodation letter'' is still present, but the tone is ``we will remove barriers,'' not ``you must prove eligibility before we consider changes.''

\subsubsection{Burden-Framed Example (community/technical college context)}
\label{sec:burden-example}

\begin{quote}\itshape
If you are a qualified student with a documented disability, it is your responsibility to self-identify 
and submit appropriate documentation to the Office of Student Services. 
The College reserves the right to determine what accommodations are reasonable. 
No accommodations will be granted until documentation is approved. 
Accommodations are not retroactive.
\end{quote}

We coded this passage as \textbf{burden-framed} because (1) all action is on the student (``your responsibility to self-identify''), (2) the institution ``reserves the right'' to judge what counts as reasonable, which signals conditional access, and (3) it explicitly withholds accommodations until approval, and denies retroactive coverage. This shifts both bureaucratic load and risk (for example, around safety procedures) onto the disabled student.

\subsection{Safety-Critical Instructions as Screen-Readable Text}
\label{sec:coding-examples:safety}

\subsubsection{Example: Welding / Shop Safety (technical college context)}

\begin{quote}\itshape
Personal Protective Equipment (PPE) is required at all times while operating GTAW equipment. 
Students must: 
(1) Inspect cables, torches, and shielding gas connections before activation; 
(2) Wear ANSI-rated eye protection and gloves rated for high-temperature work; 
(3) Maintain a clear work area free of flammable materials within a 3-foot radius. 
Failure to follow these procedures may result in removal from lab.
\end{quote}

In the source document, this guidance appeared as bullet-point \textit{text}, not an embedded image. We coded this as \textbf{safety-critical content available to screen readers}. This matters because it allows a blind or low-vision student to independently review the procedures that directly affect physical safety in a high-risk environment (hot metal, live current, gas).

\subsubsection{Example: Biomedical / Instrumentation Lab Safety (public R1 context)}

\begin{quote}\itshape
This course involves hands-on work with biomedical instrumentation and clinical-grade sensors. 
All students must follow sterile handling procedures and electrical safety guidelines. 
You are responsible for knowing the lab rules described in this syllabus and any updates announced in lecture. 
Improper handling of equipment will result in removal from the lab.
\end{quote}

Again, this appeared in body text (not as a scanned PDF image). We coded this as \textbf{safety-critical content available to screen readers}. We also flagged the phrase ``You are responsible for knowing the lab rules described in this syllabus'' as increasing the stakes: if the syllabus were not machine-readable, a blind or low-vision student could be held accountable for safety rules they could not independently access.

\subsection{Governance Model: Instructor Voice vs.\ Master Syllabus}
\label{sec:coding-examples:governance}

\subsubsection{Instructor-Authored Syllabus (R1 context)}

\begin{quote}\itshape
Instructor: Dr.\ [Name]. Office Hours: Tue/Thu 11:00--12:30 in ECJ 2.112. 
Final Exam: Thursday Dec.\ 12, 3:30--5:30pm (no exceptions). 
Project teams will design a device addressing a real clinical need. 
All course materials, including safety guidelines and accommodation procedures, 
are provided in this syllabus and on the course site.
\end{quote}

We coded this as \textbf{instructor-authored}: it includes instructor name, office hours, room numbers, and a single scheduled exam slot. Accessibility and safety expectations are embedded in the instructor's local voice.

\subsubsection{Centralized Master Syllabus (community / technical college context)}

\begin{quote}\itshape
MASTER SYLLABUS \\
Course Prefix and Number: WLD 1234 \\
Course Title: Gas Tungsten Arc Welding (GTAW) \\
Credit Hours: 4 \\
Course Description: This course provides instruction in GTAW safety procedures, equipment setup, and weld bead quality. \\
Disability Code: Qualified students with disabilities are required to self-identify and provide documentation to Student Affairs. The College reserves the right to determine appropriate accommodations. \\
Learning Outcomes (all sections): Safely set up GTAW equipment; Produce fillet welds meeting specified tolerance; Demonstrate compliance with shop safety procedures.
\end{quote}

We coded this as a \textbf{centralized master syllabus}: it is labeled ``MASTER SYLLABUS,'' includes fixed learning outcomes, a fixed disability statement, and safety language that applies to \emph{all} sections of the course. Individual instructors are expected to ``build on'' this template rather than rewrite core policy.

\subsection{Proactive Universal Design Language}
\label{sec:coding-examples:udl}

\subsubsection{Example: Universal Design Commitment (public R1 / UC context)}

\begin{quote}\itshape
Whenever possible we will use universal design practices to make course materials inclusive for everyone.
For example, slides and handouts will avoid color combinations that are difficult to distinguish for low-vision readers,
and video content will include captions. 
Students with an Authorization for Accommodation (AFA) letter should share it with the instructor 
so that individual needs can also be addressed.
\end{quote}

We coded this as \textbf{proactive universal design} because the syllabus promises inclusive formatting (captioning, color contrast) for all students by default. The accommodation letter is still required for individualized adjustments, but the baseline commitment is ``we design access in,'' not ``we will consider changes after you qualify.''

\subsubsection{Example: Reactive/Procedural Accommodation-Only Language (community college context)}

\begin{quote}\itshape
Students requesting accommodations must first obtain official approval from Disability Services.
Instructors are not permitted to modify course materials, deadlines, or safety procedures 
until documentation has been received and verified.
\end{quote}

We coded this as \textbf{reactive / procedural only}. There is no promise of proactive accessible design. Instead, the syllabus frames any flexibility (including potentially life-safety flexibility in a lab) as something the instructor is \emph{not allowed} to provide until after paperwork is processed.

The coded examples above operationalize the core claims we make later in the paper. The accommodation statements in Sections~\ref{sec:rights-example} and~\ref{sec:burden-example} illustrate two distinct rhetorical postures toward disability access. In the rights-framed case, the institution positions access as a shared obligation to remove barriers and invites the student into a collaborative process. In the burden-framed case, the institution positions access as conditional, emphasizes documentation and institutional discretion, and withholds flexibility until approval. These contrasts ground our claim in Section~\ref{sec:results} that accommodation framing is systematically different across institution types, especially between R1 universities and workforce-oriented colleges.

The safety-critical excerpts in Section~\ref{sec:coding-examples:safety} show that, in many cases, lab and shop procedures are published as selectable text rather than as screenshots or low-resolution scans. We use this to argue in Section~\ref{sec:results} that blind and low-vision students are often given screen-reader access to essential expectations about laboratory behavior, personal protective equipment, and hazardous equipment handling. At the same time, these same syllabi sometimes insist that no adjustments to procedure will be considered until disability documentation is processed (Section~\ref{sec:burden-example}). We interpret this tension in Section~\ref{sec:discussion} as a safety risk: a student may be accountable for following safety rules that they can read, but may not yet be allowed to request alternative formats, assistive tools, or modified assessment while waiting for approval.

The governance contrast in Section~\ref{sec:coding-examples:governance} demonstrates that accessibility language is not just a matter of individual instructor style. In community and technical colleges, the ``MASTER SYLLABUS'' functions as infrastructure. Its accommodation language, safety rules, and learning outcomes are replicated across every section of the course and sometimes across multiple campuses. We argue in Section~\ref{sec:discussion} that this centralized template is a high-leverage intervention point for accessibility: a rights-based, proactively inclusive master syllabus can scale equity, while a burden-framed master syllabus can scale gatekeeping.

Finally, the universal design language in Section~\ref{sec:coding-examples:udl} gives us evidence that some institutions and instructors are beginning to normalize proactive accessibility practices (for example, accessible color contrast and captioning) as default course design rather than as special treatment. We use this observation to support our claim that accessibility in higher education is not only about PDF tagging or screen reader compatibility. It is also about whether the institution frames inclusion as part of baseline instructional quality. Together, these coded patterns connect document form, safety, policy rhetoric, and institutional power, which is the core analytic move of this paper.

\begin{figure}[t]
  \centering
  \includegraphics[width=0.6\linewidth]{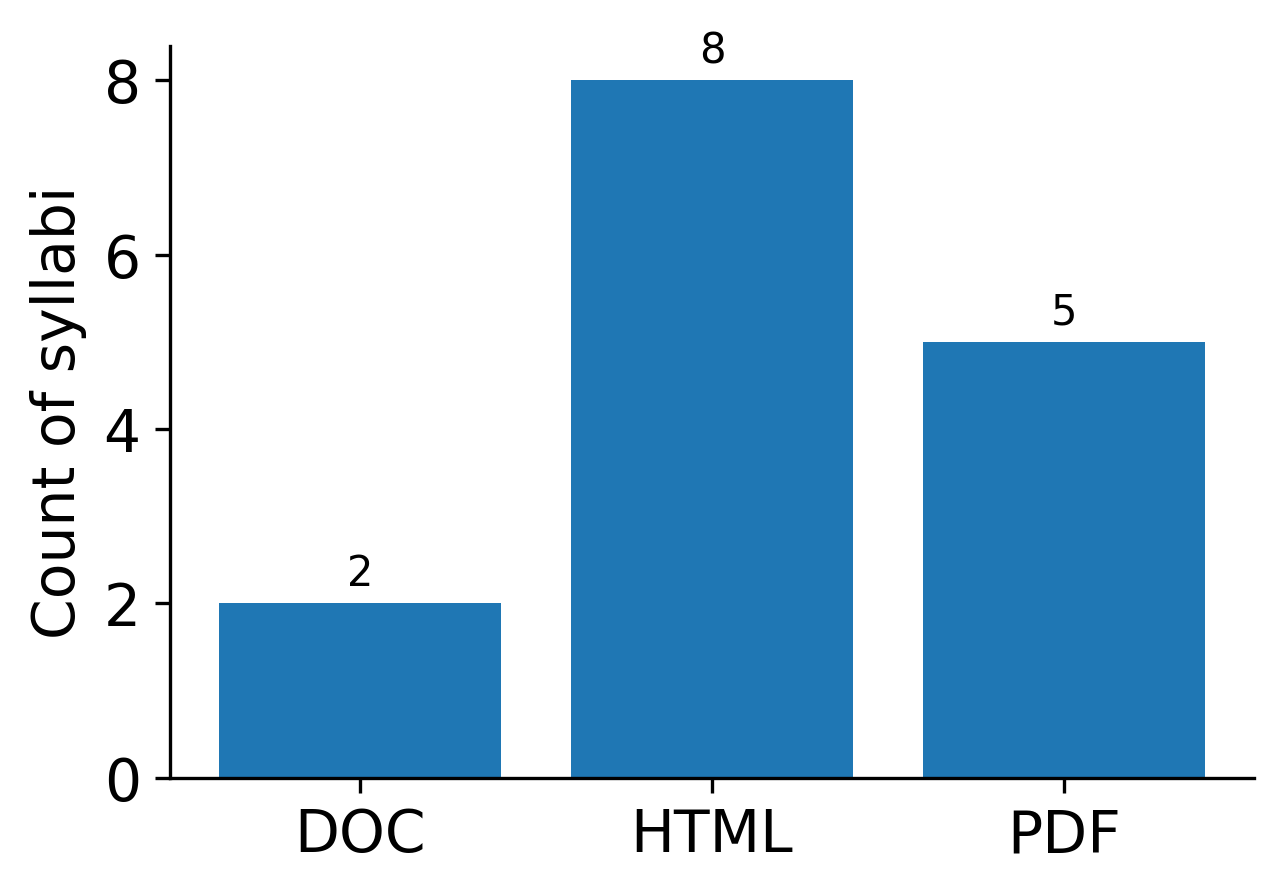}
  \caption{File-type mix in the sampled syllabi (N=15). Bars show counts of HTML pages (8), PDF documents (5), and Word documents (2). This provides context for downstream accessibility checks: HTML pages are typically machine-readable by default, whereas PDFs and DOC files may vary in structure and tagging quality.}
  \Description{A small bar chart with three bars labeled HTML, PDF, and DOC showing counts 8, 5, and 2 respectively.}
  \label{fig:filetypes}
\end{figure}

\end{document}